\documentstyle[12pt]{article}

\newcommand{\ben}{\begin{equation}}
\newcommand{\een}{\end{equation}}
\newcommand{\epsn}{\epsilon}
\newcommand{\taubar}{\bar{\tau}}
\newcommand{\betabar}{\bar{\beta}}
\newcommand{\ubar}{\bar{u}}
\newcommand{\ubarx}{\bar{u}_x}
\newcommand{\ubary}{\bar{u}_y}
\newcommand{\ubarxt}{\bar{u}_{x}(\bar{\tau})}
\newcommand{\ubaryt}{\bar{u}_{y}(\bar{\tau})}

\begin{document}

\baselineskip 22pt

\begin{center}
{\Large \bf Quantum vortex creep\\
:Hall and dissipative tunneling}\\
\vspace{1.0cm}
Gwang-Hee Kim$^1$ and Mincheol Shin$^2$\\
{\it{${}^{1}$Department of Physics, Sejong University,
Seoul 143-747, KOREA\\
${}^{2}$Research Department, Electronics and
Telecommunications\\Research Institute, Taejon 305-600, KOREA}}\\
\vspace{2.0cm}
{\bf Abstract}\\
\end{center}

Within the framework of the path-integral approach we study the
quantum vortex creep for the situation where both the Hall and the
dissipative dynamics are simultaneously present.  We calculate the
relaxation rate and the crossover temperature separating the thermal
activation and the quantum tunneling processes for anisotropic or multilayer
superconductors. The results are compared with the available
experimental data.
\\
\vfill

\noindent
PACS number(s): 74.60.Ge
\vspace{1.0cm}

\noindent
$1$: e-mail: gkim@kunja.sejong.ac.kr\\
$2$: e-mail: mcshin@idea.etri.re.kr

\thispagestyle{empty}
\pagebreak

\baselineskip 22pt

After the discovery of high temperature superconductors (HTSC) the
relaxation of magnetization has been the subject of intensive studies
because of the possibility of quantum tunneling of vortices in the
system.\cite{mot}-\cite{ao} According to the Anderson-Kim
model,\cite{and} the vortices can move out of the pinning sites by the
thermal activation process, which proceeds at a rate proportional to
$\exp (-U_0 /k_B T)$, where $U_0$ is the height of the energy barrier
generated by the pinning mechanism.  This process induces a
redistribution of vortices, hence of the current loops related with
the vortices, causing a change in the magnetization logarithmically
with time
\ben
{\bf M}(t)={\bf M_0}[1-Q(T) \ln (t/t_0)],
\een
where $Q(T)=k_B T/U_0$ is the so called magnetic viscosity.\cite{yes}
Since the rate is expected to vanish at zero
temperature, the magnetization tends to be unchanged as the
temperature $T$ approaches zero.  However, it has been realized that
the thermal activation theory fails to explain experimental
observations of temperature independent magnetic relaxation in
HTSC\cite{mot} at low temperatures. This has led to introduce the
concept of quantum tunneling of vortices trapped in the pinning
potential with the collective pinning length $L_c$.\cite{bla} Since a
pinned vortex is in some metastable state, the quantum tunneling of
the vortex is governed by the frequency of the pinning barrier
hindering the tunneling process. The rate of the quantum tunneling can
be thus obtained with the relevant dynamics, whereas the rate of the
thermal activation process does not depend on particular dynamics.

Recently, Blatter et al.\cite{bla1} studied the quantum vortex
tunneling for the case where the dissipative term is dominant in the
dynamical equation of motion of vortex, and many experimental
results\cite{mot} have been interpreted within this limit. On the
one hand, Feigel'man et al.\cite{fei} estimated the low-lying level
spacing ($\sim \hbar \omega_0 \simeq \Delta^2/ \epsn_F$) in the vortex
core and the transport relaxation time of the charge 
carriers $\tau_r$ at $T=0$ by using a linear
extrapolation of the normal state resistivity $\rho_n$,
and obtained $\omega_0 \tau_r \gg 1$ in HTSC
at low temperature which is the superclean limit. The
criteria distinguishing the dissipative and the Hall vortex motion is
the magnitude of the quantity $\omega_0
\tau_r$,\cite{kop} therefore they proposed that the Hall motion should
be dominant in clean HTSC. Most recently, however, van Dalen et
al.\cite{dal}-\cite{dal2} observed experimentally that in YBCO or
BiSCCO systems, the vortex tunneling at low temperatures may occur in
an intermediate regime between the purely dissipative and the
superclean Hall tunneling. Prior to the experiments, Stephen\cite{ste}
studied the quantum tunneling of vortex with the dissipative and the
Hall terms by using the approximate form for the mean-square quantum
fluctuation with a simple harmonic pinning potential up to a
cutoff. As is noted in Ref.~\cite{cal}, since the cutoff model
potential can be considered to be of the form $(q/q_0)^2 - (q/q_0)^n$
with $n \rightarrow \infty$ where $q$'s are coordinates of the
potential, the steep part of the model potential underestimates the
suppression factor, i.e., the effective action in the case of damped
systems. So they remarked that a special care should be taken in
replacing the smooth potential by one which has a discontinuity.

In this paper, we study the quantum vortex tunneling for the case
where the dissipative and the Hall dynamics are simultaneously
present, considering the pinning potential barrier 
generated by disorder with a cubic
term\cite{bla,fei,chu} along the $x$ axis near criticality, which
should better model the real pinning potential.  Since the
tunneling rate takes the form $\Gamma_{Q} \propto \exp(-S_{\rm
cl}/\hbar)$, where $S_{\rm cl}$ is the least action obtained from the
classical trajectory, we calculate the classical equations of motion
in the imaginary time path-integral formalism.  When both the dynamics are
simultaneously considered for the equation of motion, the resultant
integral equation becomes complicated by the presence of the cubic
term in the pinning potential, which compels us to rely on numerical
methods to obtain the solutions.

Our considerations will begin by discussing the pancake
vortex in the $xy$ plane with the length $L_c$ along the $z$ axis, where
$L_c$ is the collective pinning length which can be expressed in terms
of the mass anisotropy parameter $\varepsilon^{2}_{a}=m/M < 1$, the
coherence length $\xi$, and the depairing and critical current
densities $j_0$ and $j_c$; $L_c \simeq \varepsilon_a 
\xi (j_0 /j_c)^{1/2}$ within the weak collective pinning
theory.\cite{bla} The supercurrent characterized by the parameter
$\epsn=(1-j/j_c)^{1/2}$ is along the $y$ direction.
Since the length scale in this study is much larger than the size of a
vortex core, a vortex can be regarded as a point-like object.
Neglecting the inertial mass term in the equation of motion,
Kopnin et al.\cite{kop} derived the following equation of motion:
\ben
\eta {\bf \dot{u}} + \alpha {\bf \dot{u}} \times \hat{z}=
{\phi_0 \over c} {\bf j} \times \hat{z} + {\bf F}_p,
\label{eom}
\een
where $\phi_0=h c /2e$ is the flux quantum,
${\bf u}$ the distortion
of the vortex from the equilibrium position in the $xy$ plane,
${\bf F}_p$ the pinning force, and $\eta$ and $\alpha$ are
the viscous and Hall drag coefficients given by\cite{bla,kop}
\ben
\alpha=\alpha_0 {(\omega_0 \tau_r)^2 \over
        1+(\omega_0 \tau_r)^2}, \ \ \
\eta=\alpha_0 {\omega_0 \tau_r \over 1+(\omega_0 \tau_r)^2}.
\een
Here $\alpha_0=\pi \hbar n_s$,
$\tau_r=m/n e^2 \rho_n$, $\Delta=\hbar v_F/\pi \xi$,
and $n_s$ is the density of charge carriers,
$m$ its effective mass, and
$v_F$ the Fermi velocity.     

To study the tunneling process at zero temperature, 
we need to consider the imaginary time path integral\cite{lan}
\ben
\int D{u_x} \int D{u_y} \exp (-S_E/\hbar).
\een
Considering the effect of dissipation on the tunneling, we use
the model introduced by Caldeira and Leggett\cite{cal} 
and others\cite{bla,fei,ste}
in which dissipation is modeled by
coupling the vortex degree of freedom to a bath of harmonic oscillators.
By combining the pinning potential with a
cubic term with the Lorentz term in Eq. (\ref{eom}) 
the model potential for the pancake vortex 
is taken as\cite{bla,fei,chu}
\ben
V_p (u_x, u_y)- {\phi_0 j u_x \over c} \equiv
{V_0 \over 2}[({u_y \over R})^2+
c_{1}\epsn ({u_x \over R})^2-
{2\over 3} c_{2} ({u_x \over R})^3],
\label{pinpotappx}
\een
where $\epsn=\sqrt{1-j/j_c}$, $c_{1,2} \sim 1$,  
$V_0 \sim (\phi_0 / 4 \pi \lambda_{xy})^2$, and
$\lambda_{xy}$ is the bulk planar penetration depth.\cite{com1}

Now, the Euclidean action of this system takes 
the form\cite{wei}
\begin{eqnarray}
 S_E &=& L_c \int_{-\infty}^{\infty} d\tau
     \{ -i\alpha {du_x \over d\tau} u_y +
     {V_0 \over 2}[({u_y \over R})^2+
     c_{1}\epsn ({u_x \over R})^2-
     {2\over 3} c_{2} ({u_x \over R})^3]
     \nonumber \\
    & & + {1 \over 2}
     \int_{-\infty}^{\infty}d\tau_1
     K_0(\tau-\tau_1)[{\bf u}(\tau)-{\bf u}(\tau_1)]^2 \},
\label{euact1}
\end{eqnarray}
where the so called nonlocal influential function
becomes
$K_0(\tau)={1 \over 2 \pi} \int_{0}^{\infty} d\omega
J(\omega) \exp(-\omega |\tau|)$ where
$J(\omega)$ is the spectral density whose form will
be discussed later. 

Let us now introduce $u_{x0}$, $u_{y0}$, and $\tau_0$ as the length
scales for $u_x$, $u_y$, and $\tau$, respectively, which are to be set
by the following scaling analysis for Eq.\ (\ref{euact1}).\cite{mor}
A comparison
of the quadratic potential term with the cubic one in the $x$
direction determines the relevant length scale along $x$; $u_{x0} \sim
\epsn R c_1 /c_2$. Equating the quadratic potential terms in
the $x$ and $y$ directions, we obtain the length scale for the displacement
in the $y$ direction; $u_{y0} \sim \epsn^{3/2} R c^{3/2}_{1}/
c_2$. The time scale for the imaginary time, $\tau_0
\sim \alpha R^2/(V_0 \sqrt{c_1 \epsn})$, is set by comparing the
Hall term with the quadratic potential in the $y$ direction.
Keeping these points in mind, it is convenient to use 
the dimensionless variables
\ben
u_x=({c_1 \epsn R \over 2 c_2}) \bar{u}_x, \ \ \
u_y=({c_{1}^{3/2} \epsn^{3/2} R \over 2 c_2})
\bar{u}_y, \ \ \
\tau=({\sqrt{2} R^2 \alpha_0 \over
\sqrt{c_1 \epsn} V_0}) \bar{\tau}.
\een
The Ohmic dissipation\cite{bla,fei,ste,mor} in Eq. (\ref{eom}),
which is obtained if  
$J(\omega)=\eta \omega$ after elimination of the oscillators,
leads to the influential function
\ben
K_0(\tau)=\eta /2 \pi |\tau|^2.
\een
Then, the Euclidean action is simplified as
\begin{eqnarray}
 S_E &=&
      ({\sqrt{2} c_{1}^{5/2}\over 4 c_{2}^{2}})
      (L_c \alpha_0 R^2)
      \epsn^{5/2}  I,
      \label{se}\\
 I&=&\int_{-\infty}^{\infty} d\bar{\tau}
     \{ -i\alpha_1 {d\ubar_x \over d\bar{\tau}}
     \ubar_y +
     {1 \over 2}\ubar_{y}^{2}+
     {1 \over 2}\ubar_{x}^{2}-{1 \over 6}\ubar_{x}^{3}
     \nonumber \\
    & & + {1 \over 4} \eta_1
     \int_{-\infty}^{\infty}d\bar{\tau}_1
     {[\ubar_x(\taubar)-\ubar_x(\taubar_1)]^2+
     c_1 \epsn [\ubar_y(\taubar)-\ubar_y(\taubar_1)]^2
     \over |\taubar-\taubar_1|^2} \},
\label{IHD}
\end{eqnarray}
where the dimensionless Hall and dissipation coefficient 
are given by
\ben
\alpha_1= {1 \over \sqrt{2}}
{(\omega_0 \tau_r)^2 \over 1+(\omega_0 \tau_r)^2}, \ \ \
\eta_1={\sqrt{2} \over 2 \pi \sqrt{c_1 \epsn}}
 {\omega_0 \tau_r \over 1+(\omega_0 \tau_r)^2}.
\label{eta1}
\een

Let us briefly consider the case of a pure Hall type 
which is relevant in super-clean superconductors.\cite{fei} 
In this limit the integral (\ref{IHD}) does not include $\eta_1$-term.
Noting that $I \sim \alpha_1 \int_{-\infty}^{\infty} d\bar{\tau}
\bar{u}^{\rm cl}_{y} d\bar{u}^{\rm cl}_{x}/d\bar{\tau}
=\alpha_1 \times$
(numerical quantity) where $\bar{u}^{\rm cl}_{x}$ and
$\bar{u}^{\rm cl}_{y}$ are classical trajectories in the
Hall limit, the approximate least action takes the simple form
\ben
S_{\rm cl} \sim (L_c \alpha_0 R^2) \epsn^{5/2} \alpha_1
\sim (L_c \alpha_0 R^2) \epsn^{5/2} ,
\een
which agrees up to a numerical factor with the result in
Table \ref{tabsum} obtained by the explicit instanton solution.
In the case where the dissipation term is dominant in the vortex
dynamics,\cite{bla} $\alpha_1$-term is not included in 
the integral (\ref{IHD}). Noting that $I \sim \eta_1 \times$
(numerical quantity) and $\eta_1 \sim \omega_0 \tau_r /\sqrt{\epsn}$ 
in this limit, $S_{\rm cl}$ is approximately
\ben
S_{\rm cl} \sim (L_c \alpha_0 R^2) \epsn^{5/2} \eta_1 
\sim (L_c \alpha_0 R^2) \epsn^2 (\omega_0 \tau_r),
\label{sdiss}
\een
which is also consistent with the result in Table \ref{tabsum} up to 
a numerical factor.\cite{com1-1}

In the intermediate regime, i.e. 
when the Hall and the dissipative dynamics are
simultaneously present,  the classical trajectories of $\ubarx$ and
$\ubary$ should be determined by the Euler-Lagrange equation derived
from Eq. (\ref{IHD}):
\begin{eqnarray}
 i\alpha_1 {d \ubary \over d\taubar}+\ubarx
-{\bar{u}_{x}^{2} \over 2}
 -\eta_1 \int_{-\infty}^{\infty}d\taubar_1 
 ({d\ubarx \over d \taubar_1}) 
{1 \over \taubar_1-\taubar}&=&0,
 \label{eq:h-d-t1}\\
 -i\alpha_1 {d\ubarx \over d\taubar}+\ubary
 -\eta_1 c_1 \epsilon \int_{-\infty}^{\infty}d\tau_1
 ({d\ubary \over d \taubar_1})
{1 \over \taubar_1-\taubar}&=&0,
 \label{eq:h-d-t2}
\end{eqnarray}
where $\ubarx(-\taubar)=\ubarx(\taubar)$
and $\ubary(-\taubar)=-\ubary(\taubar)$.
As in the dissipative case,
the equations (\ref{eq:h-d-t1}) and (\ref{eq:h-d-t2}) are
nonlocal in time, which necessiates
the Fourier transformation as follows:
\begin{eqnarray}
 [\frac{(\alpha_1 \omega)^2}{1+\pi \eta_1
 c_1 \epsn |\omega| }
 &+&1+\pi \eta_1 |\omega| ] \ubarx (\omega)
 \nonumber \\
 &=&\frac{1}{\sqrt{8 \pi}} \int_{-\infty}^{\infty}
 d\omega_1 \ubarx (\omega - \omega_1) \ubarx (\omega_1),
 \label{eq:h-d-w1}
\end{eqnarray}
and
\begin{equation}
 \ubary (\omega)={-\alpha_1 \omega \over 1+\pi \eta_1
 c_1 \epsilon |\omega| } \ubarx(\omega)
 \label{eq:h-d-w2}
\end{equation}
where $\ubarx(-\omega)=\ubarx(\omega)$ and
$\ubary(-\omega)=-\ubary(\omega)$. Also, we note that
Eq. (\ref{IHD}) is reduced to 
\begin{eqnarray}
I&=&\int_{-\infty}^{\infty} d\omega 
\{ {1 \over 2} [\frac{(\alpha_1 \omega)^2}{1+\pi \eta_1
 c_1 \epsn |\omega| }
 +1+\pi \eta_1 |\omega| ] \ubarx (\omega) \ubarx (-\omega)
\nonumber\\
& & - {1 \over 6} \int_{-\infty}^{\infty} {d \omega_1 \over \sqrt{2 \pi}}
\ubarx (\omega) \ubarx (\omega_1) \ubarx (\omega+\omega_1) \},
\label{sw}
\end{eqnarray}
after Fourier transform and integrating over the $\ubary(\omega)$
variable.
 
Even though Eq. (\ref{eq:h-d-w1}) is a one-dimensional problem
with respect to $\ubarx$, its analytic solution 
becomes complex by the presence of
the nonlocal term arising from the cubic
potential (\ref{pinpotappx}). We have
therefore numerically solved the above integral equations
as follows.

Let us introduce, for a given function $\cal F = \cal F(\omega)$,
\begin{eqnarray}
\cal L_{\cal F}(\omega) &\equiv& \left\{{(\alpha_1 \omega)^2 \over
1+\pi\eta_1 c_1 \epsilon|\omega|} + 1 +
\pi\eta_1|\omega|\right\}\cal F(\omega),\\
\label{eq:LF}
\cal R_{\cal F}(\omega) &\equiv& {1\over
\sqrt{8\pi}}\int_{-\infty}^{\infty}d\omega_1
{\cal F}(\omega-\omega_1)
{\cal F}(\omega_1).
\label{eq:RF}
\end{eqnarray}
With the function $\cal F(\omega)$ in the above definitions replaced
by the Fourier transformed classical path $\ubarx(\omega)$, 
the equation (\ref{eq:h-d-w1}) corresponds to  
\begin{equation}
{\cal L_{\cal F}(\omega) - \cal R_{\cal F}(\omega)} = 0.
\label{eq:LR}
\end{equation}
Let us now define, for a
given function $\cal F$,
\begin{equation}
\label{eq:delta}
\delta_{\cal F} \equiv \lim_{\Omega\rightarrow\infty}{1\over \Omega}
\int_{0}^{\Omega}d\omega \left|\cal L_{\cal F}(\omega) -
\cal R_{\cal F}(\omega)\right|.
\end{equation}
If a trial function ${\cal F}$ has a set of adjustable parameters $\{p_i\}$,
$\delta_{\cal F}$ becomes a function of $\{p_i\}$ and if the set
$\{p^*_i\}$ which minimizes $\delta$ can be found, the function ${\cal
F}$ with the parameter set $\{p^*_i\}$ will be the desired solution of
the integral equation. As shown in Table \ref{tabsum}, a natural
choice for the trial function $\cal F$ for general values of
$\alpha_1$ and $\eta_1$ would be to linearly combine the two extreme
solutions such as
\begin{equation}
\label{eq:func_trial}
{\cal F}(\omega) = p_1 \omega/\sinh(p_2\omega) + p_3
\exp(p_4|\omega|),
\end{equation}
where $p_1$, $p_2$, $p_3$, and $p_4$ are parameters to be controlled
in the variational procedure. 

We have carried out minimization of $\delta$ of Eq.\ (\ref{eq:delta})
with the trial function of Eq.\ (\ref{eq:func_trial}) by the conjugate
gradient method. It turns out that the variational method is very
successful, covering all ranges of $\omega_0\tau_r$ for arbitrary
values of $\epsilon$. In principle, a perfect minimization of $\delta$
should make it to be zero, but due to the numerical accuracy in
carrying out integrations involved in the procedure (or it may be
possible that the trial function of the form in
Eq.\ (\ref{eq:func_trial}) cannot perfectly minimize $\delta$), our
numerical calculation typically gave $\delta/{\bar{\cal L}} \leq
10^{-3}$, where ${\bar{\cal L}}$ is the average value of ${\cal
L_{\cal F}}(\omega)$ in the integrand of
Eq.\ (\ref{eq:delta}). Nevertheless, the accuracy is far good enough
for our purpose of evaluation of the action integral in later
stages. The accuracy of the variational method has been also checked
by comparing the solutions by this method with those by the direct
iterations of differential equations (\ref{eq:h-d-t1}) and
(\ref{eq:h-d-t2}) in the range of $\omega_0\tau_r$ where the solutions
by the iteration method are available.\cite{com2} The two solutions
always match excellently, which confirms that our variational method
produces true instanton solutions.

Typical instanton solutions thus obtained are illustrated in
Fig.\ \ref{figinst}. The classical Euclidean action $S_{\rm cl}$ is given
by a numerical integration of Eq.\ (\ref{IHD}) with the instanton
solutions. We then have the relaxation rate or the magnetic viscosity
$Q(T=0)$ which equals $\hbar/S_{\rm cl}$, in other words, $Q(0)/Q_0= 2
\sqrt{2}/I$ where $Q(0) \equiv Q(T=0)$ and $Q^{-1}_{0}=\epsn^{5/2}(\pi
n_s L_c\xi^2)$. Here we have set $c_{1,2}$ to be 1 and replaced
$R$ by the coherence length $\xi$. The results of evaluation of
$Q(0)/Q_0$ versus $\omega_0 \tau_r$ for $\epsn=1,0.1,0.01$, and
$0.001$ are shown in Fig.\ \ref{figQ}. We first point out that our
$Q(0)/Q_0$ is different from the result in
Ref.~ \cite{ste}, which reflects the fact that the
infinitely steep potential which underestimates the suppression factor
is carefully treated in our model potential.\cite{com4} As is noted in Table
\ref{tabsum}, the approximate value of $\epsn^{5/2}Q(0)$ in units of
$\pi n_s L_c \xi^2$ is 15/18 in the Hall limit and $2\epsn^{1/2}/(\pi
\omega_0\tau_r)$ in the dissipative limit, which agree with the
limiting values of the figure. Even though $Q(0)/Q_0
(=\hbar/S_{\rm cl}$) looks going toward infinity as
$\omega_0\tau_r \rightarrow 0$, it actually  does not diverge in that
limit. In the pure dissipative case, the classical action
$S_{\rm cl}$ vanishes as $\omega_0\tau_r \rightarrow 0$ in our 
consideration, which is expected from the limiting behavior of 
Eq. (\ref{sdiss}). However, in this situation 
the inertia term not included in this work influences the results.
By including the massive term, in this limit
the approximate form of the classical action becomes
$S_{\rm cl}/\hbar \sim L_c \sqrt{m_v V_0} \xi 
\epsn^{5/2}$ which is independent of $\omega_0\tau_r$ where
$m_v$ is the inertia mass of a vortex.\cite{bla,suh}
In general the massive term is small in the region except for the
limit mentioned before and usually can be neglected as compared to
either the dissipative or the Hall term.
In the intermediate tunneling regime,
our curves display an interesting feature which is the clear existence
of minimum of $Q(0)/Q_0$ for each $\epsn$.  Even at $\epsn=1$, there
exists a shallow yet clear minimum near $\omega_0\tau_r \sim 3.5$ (see
the inset). As $\epsilon$ becomes smaller, i.e., as $j \rightarrow
j_c$, the minimum becomes much more pronounced with its location
moving toward $\omega_0\tau_r \approx 1$ at the same time.  
The existence of the minimum of $Q$ can be understood by noting
the following qualitative features of the relaxation rate based on
the dimensional estimation.\cite{bla} 
Noting that for the pure Hall motion
the classical trajectory is
dominant in the frequency scale $|\omega| \sim 1/\alpha_1$
in Eq. (\ref{eq:h-d-w1}) which is also understood by the solution
$\ubarx(\omega)$ in Table \ref{tabsum},
we find that from Eqs. (\ref{se}) and (\ref{sw})
the correction to the pure Hall classical action 
$S^{H}_{0}$ by the small dissipation $\eta_1$ is of the order
$\eta_1/\alpha_1$, that is, the corrected classical action becomes
$(1+\eta_1/\alpha_1)S^{H}_{0} \sim (1+1/\omega_0 \tau_r) S^{H}_{0}$.
The relaxation rate is then $Q \sim \omega_0 \tau_r/(1+\omega_0 \tau_r)$,
which increases monotonically as a function of $\omega_0 \tau_r$ in the
large $\omega_0 \tau_r$ regime. The fact that the classical action 
increases (or the corresponding relaxation rate decreases) by
inclusion of the dissipation is well known in the macroscopic
quantum tunneling problems, which is also 
physically clear in the sense that the dominant (Hall) motion is hindered by
the frictional force, resulting in decreasing the relaxation rate.
In the opposite limit, on the other hand, the correction to 
the pure dissipative action $S^{D}_{0}$ by the small Hall contribution
$\alpha_1$ is of the order $(\alpha_1/\eta_1)^2$. The corrected
classical action is then $(1+(\alpha_1/\eta_1)^{2}) S^{D}_{0}
\sim \omega_0 \tau_r (1+(\omega_0 \tau_r)^2 )$, leading to the 
relaxation rate $Q \sim 1/[\omega_0 \tau_r (1+(\omega_0 \tau_r)^2)]$.
$Q$ now decreases monotonically
as a function of $\omega_0 \tau_r$ in the small 
$\omega_0 \tau_r$ regime. Therefore, it is expected that there exists a 
minimum of $Q$ in the intermediate range of $\omega_0 \tau_r$ because
the relaxation rate decreases monotonically for small 
$\omega_0 \tau_r$ but increases monotonically for 
large $\omega_0 \tau_r$.
This suggests us to predict observation of minimum in an experiment.

As is shown in Fig.\ \ref{figQ}, our results are valid in any value of 
$\epsn \leq 1$. It is well known that the tunneling rates get larger
for smaller $\epsn$ since the height and the width of the barrier are
proportional to the parameter $\epsn$. However,
there might exist the quantum tunneling even in the absence of
the external current, i.e., $\epsn=1$.
Actually the existence of the quantum tunneling is determined by estimating
the magnitude of the quantum tunneling rate $\Gamma_Q$.  
If we apply our result of $\epsn=1$ to the existing experimental 
data,\cite{com1}
where $Q(0)/Q_0 \approx 2.3$  and $Q_{0}^{-1} \approx 78$
in the $\rm {Y Ba_{2} Cu_3 O_7}/\rm {Pr
Ba_{2} Cu_3 O_7}$ multilayer systems\cite{dal} and $Q(0)/Q_0 \approx
2.0$ and $Q_{0}^{-1} \approx 66$ 
in the $\rm {Bi Sr_{2} Ca Cu_2 O_8}$ single crystal,\cite{dal1}
we obtain $S_{\rm cl}/\hbar \approx 33$ for both systems.
This gives $\Gamma_{Q} \approx \omega \exp(-S_{\rm cl}/\hbar)
\approx 4.66 \times 10^{-3}\ (4.66 \times 10^{-4}) {\rm sec^{-1}}$ for
the attempt frequency $\omega(\approx V_0 \epsn^{1/2}/(\xi^2 n_s))$
of order $10^{12}\ (10^{11}) {\rm sec^{-1}}$ for YBCO/PBCO\ (BiSCCO),
which is a resonable value to ensure the observable rate of tunneling.
Also, using the value of $Q(0)/Q_0$ for both systerms,
we get $\omega_0\tau_r \approx 0.29$, which corresponds to the Hall
angle $\Theta_H =\arctan(\alpha_1/\eta_1)=\arctan(\omega_0 \tau_r)
\approx 16^{\circ}$ for the former, and
$\omega_0 \tau_r \approx 0.37$ ($\Theta_H \approx 20^{\circ})$ for the
latter.
This implies that for those systems the vortex tunneling at
low temperatures occurs in an intermediate regime between the purely
dissipative tunneling and the superclean Hall tunneling.  It should be
also noted that the quantum creep in the oxygen deficient $\rm {Y Ba_2
Cu_3 O_x}$ films\cite{dal2} occurs in the intermediate regime as well,
where the Hall angle depends on the oxygen content.

In the thermal activation regime, the classical solutions 
$\ubarxt$ and $\ubaryt$ do not depend on $\bar{\tau}$\cite{ivl}
and the integration range in Eq. (\ref{IHD}) reduces to between 
$-\betabar/2$ and $\betabar/2$ where 
$\betabar=\beta \hbar(\sqrt{c_1\epsn}V_0/\sqrt{2}R^2 \alpha_0)$. 
Then we obtain the classical
action $S_{\rm cl} \equiv S_T= \beta \hbar U_0$, where
$U_0 = L_c V_0 \epsn^3/6$, and the corresponding escape rate
\ben
\Gamma_{T} \propto \exp(-S_T /\hbar)=\exp(-U_0/k_B T),
\een
which is the Boltzmann formula representing a pure thermal activation.
Comparing 
$\Gamma_T$ with $\Gamma_Q(\propto \exp(-S_{\rm cl}/\hbar))$,
we have the crossover temperature 
$k_B T_0 \sim A \epsn^{1/2} Q(0)/Q_0$ from the thermal to quantum regime
where $A=\phi^{2}_{0}/(96 \pi^3 \lambda^2 \xi^2 n_s)$ and
find that the quantum process dominates at
$ T <  T_0 $. 
Using $\xi=1.5 {\rm nm}$ ($2.0 {\rm nm})$, $\lambda=150 {\rm
nm}$ ($250 {\rm nm}$),\cite{com3} and $n_s = 5 \times 10^{27}/ {\rm
m^3}$ ($3.5 \times 10^{27}/ {\rm m^3}$) for YBCO multilayer systems
(BiSCCO single crystal), for $\epsn=1$
we get $T_0 \sim$ 9.48 K (2.36 K) which agrees
reasonably with the experimental results.

In conclusion, we have studied the quantum vortex creep for the case
where the Hall and the dissipative dynamics are simultaneously present
by using the standard instanton method. We have introduced the cubic
potential term to better model the pinning potential and solved the
resultant integral equations numerically.  In a comparison with
available experimental data, our results indicate that the quantum
vortex creep may occur in an intermediate regime between the Hall and
the dissipative regimes for highly anisotropic and multilayer
superconductors. It has been also found that there exists a minimum in
a scan of the relaxation rate by $\omega_0\tau_r$ for every $\epsilon$
we have studied, whose observation we predict in an experiment
with nonzero external currents. The temperature is set to zero in this
study, but an extension to finite temperatures will be discussed
elsewhere.  It will be also interesting to apply our approach to 
systems with columnar defects produced by irradiation with heavy ions.

This work was supported
by Research-Fund, Korea Research Foundation 1996 and by
the Ministry of Information and Communications of Korea.

\pagebreak

\pagebreak

\begin{table}
\caption{Instantons in the $\bar{\tau}$ space and in the $\omega$
space, and the corresponding actions in the Hall limit
($\omega_0 \tau_r \gg 1$) and 
the dissipative limits($\omega_0 \tau_r \ll 1$).
Here $c_H=18 c_{1}^{5/2} / 15 c_{2}^{2}$
and $c_D=\pi c_{1}^{2} /  2 c_{2}^{2}$.}
 \label{tabsum}
\end{table}

\begin{center}

\begin{tabular}{|c|c|c|}  \hline
$ $ &
{\rm Hall\protect\cite{fei}} &
{\rm dissipative\protect\cite{bla1,lar}}
\\ \hline
$\ubarx(\taubar)$ &  $3/\cosh^2(\taubar/2\alpha_1)$ &
$4/[1+(\taubar/\pi \eta_1)^2]$
\\
$\ubarx(\omega)$ &
${12 \alpha_1 \over \sqrt{2 \pi}}
{(\alpha_1 \pi \omega) \over \sinh(\alpha_1 \pi \omega)}$ &
$(2 \pi)^{3/2}  \eta_1 \exp(-\pi \eta_1 |\omega|)$
\\
$S_{\rm cl}/(L_c \alpha_0 R^2)$ &
$c_H \epsn^{5/2} 
{(\omega_0 \tau_r)^2 \over 1+(\omega_0 \tau_r)^2}$ &
$c_D  \epsn^{2}
{\omega_0 \tau_r \over 1+(\omega_0 \tau_r)^2}$
\\
\hline
\end{tabular}
\end{center}

\pagebreak

\begin{figure}
\caption{Typical instanton solutions: $\bar{u}_x(\bar{\tau})$ (top) and
$-i\bar{u}_y(\bar{\tau})$ (bottom) for $\epsilon = 1$ when $\omega_0
\tau_r \gg 1$ (a), $\omega_0
\tau_r=1$ (b), and $\omega_0 \tau_r=0.35$ (c).}
\label{figinst}
\end{figure}

\begin{figure}
\caption{Dependence of the quantum vortex
creep relaxation rate $Q(0)/Q_0$ on $\omega_0\tau_r$ at zero
temperature, for $\epsn=0.001$ (a), 0.01 (b), 0.1 (c), and 1 (d). Also
is shown, to logarithmic accuracy, the relaxation
curve (e) from Ref.~{\protect\cite{ste}} (Eqs. (16) and
(20)). Note
that the approximate value of $Q(0)/Q_0$ becomes 15/18 in the Hall
limit ($\omega_0 \tau_r \gg 1$) and $2\epsn^{1/2}/(\pi\omega_0
\tau_r)$ in the dissipative limit ($\omega_0\tau_r \ll 1$).
In the inset the plot (a) for $\epsilon = 1$ is magnified.}
\label{figQ}
\end{figure}


\begin{thebibliography}{99}
\bibitem{com1} Even though the potential in 
Eq.\ (\ref{euact1}) is widely used in the limit of $j \rightarrow
j_c$,  the quadratic-plus-quartic potential form can be taken even for
the case $j=0$, as is the case in Ref. \cite{bla,fei}.
\bibitem{com1-1}
For the case of a flux line the length scale $L_c$ along the line
is inversely proportional to $\sqrt{\epsn}$, which is
simply determined by the comparision of the quadratic potential
in the $u_x$ direction with the elastic term 
along the $z$-direction.\cite{mor}
Then we obtain $S_{\rm cl} \sim \epsn^2$ for the pure Hall case
and $\epsn^{3/2}$ for the dissipative one where we used 
$\eta_1 \sim \epsn^{-1/2}$ from Eq. (\ref{eta1}).
\bibitem{com2} The iteration scheme sensitively 
depends on the choice of the initial trial functions. The most severe
disadvantage of the method is that when 
$\alpha_1/\eta_1(=\omega_0 \tau_r \pi \protect\sqrt{c_1 \epsn})$ 
becomes
smaller than $\sim 0.3$, iterations converge very slowly, and in worse
cases, even if we choose the initial trial functions well
approximating the actual solutions, slight numerical errors can drive
iterations away from convergence so as to make the iteration scheme to
break down.
\bibitem{com4} 
In Eq.\ (20) of Ref.\ \cite{ste}, the coefficient($\omega_c/\omega_{\rm
cl}$) of the linear term ($x$, i.e. $1/\omega_0 \tau_r$) ambiguously
alters the magnitude of the relaxation rate. The existence of the
minimum in the relaxation rate curve generated by the equation, which
can be seen for some values of the coefficient, also become ambiguous
due to the parametric nature of the coefficient.
\bibitem{com3} The penetration depth $\lambda$ is not given in
Ref.\ \cite{dal1}. So, we used the value for
$\rm {Bi Sr_{2} Ca Cu_2 O_8}$ in Ref.\ \cite{pol}.
\end{thebibliography}
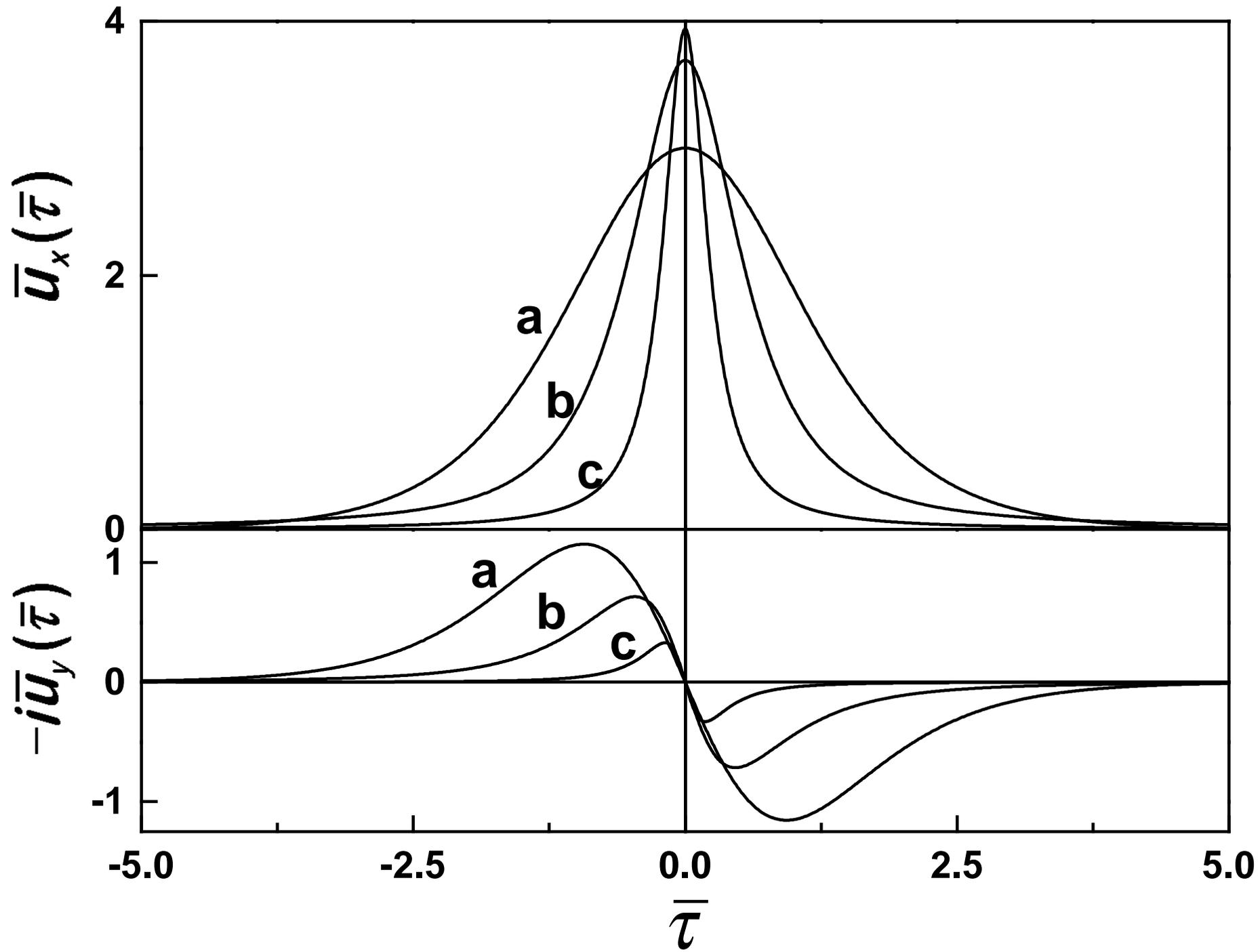
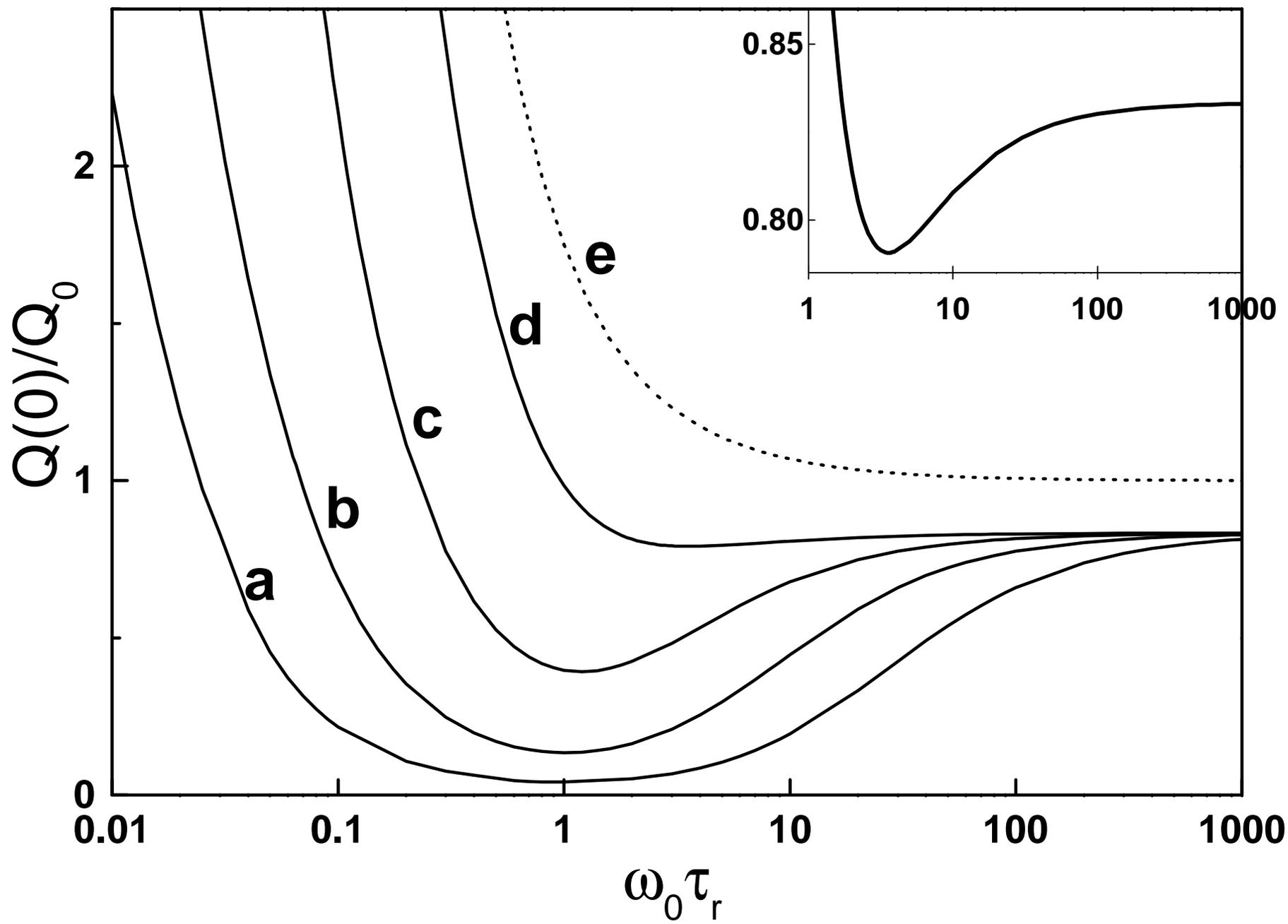
\end{document}